# Real-Time FPGA-based Digital Signal Processing and Correction for a Small Animal PET

Jiaming Lu, Lei Zhao, *Member, IEEE*, Kairen Chen, Peipei Deng, Bowen Li, Shubin Liu, *Member, IEEE*, and Qi An, *Member, IEEE*

*Abstract*—Small animal Positron Emission Tomography (PET) is dedicated to small animal imaging, which requires high position and energy precision, as well as good flexibility and efficiency of the electronics. This paper presents the design of a digital signal processing logic for a marmoset brain PET system based on LYSO crystal arrays, SiPMs, and the resistive network readout method. We implement 32-channel signal processing in a single Xilinx Artix-7 Field-Programmable Gate Array (FPGA). The logic is designed to support four online modes which are regular data processing mode, flood map construction mode, energy spectrum construction mode, and raw data mode. Several functions are integrated, including two-dimensional (2D) raw position calculation, crystal locating, events filtering, and synchronization detection. Furthermore, a series of online corrections is also integrated, such as photon peak correction to 511 keV and time measurement result correction with crystal granularity. A Gigabit Ethernet interface is utilized for data transfer, Look-Up Tables (LUTs) configuration, and command issuing. The pipeline logic works at 125 MHz with a signal processing capability beyond the required data rate of 1,000,000 events/s/channel. A series of initial tests are conducted. The results indicate that the logic design meets the application requirement.

## I. Introduction

POSITRON Emission Tomography (PET) [1][2] is a noninvasive medical imaging technique used to evaluate the metabolism level, biochemical reaction, and functional activity of various organs quantitatively and dynamically. In the field of biomedicine, studies need to be completed on small animals before extrapolation toward the humans, which requires high-quality small animal PET systems [3][4]. High demands on the performance of their readout electronics include high resolution of position, time, and energy measurement, real-time signal processing ability, as well as compatibility for multiple working modes.

In this domain, efforts have been devoted to research signal processing, system calibration and correction method. For example, in the individual readout method and row-column summing readout method for PET, a favorable way is to employ FPGAs to calculate the measurement results in real time [5][6][7]. The PET system that our electronics are used for is based on another one of the mainstream methods, named resistive network readout approach [8]. As for this method, it is required to calculate the position information from the four corners of the resistive network and locate the corresponding crystal address in an array through Look-Up Tables (LUTs), and besides there exists mismatch among these crystal cells both for energy and time measurement, while the calibration and correction process is also quite complicated. Therefore, the digital signal processing and correction for this method is more complex, so it is important to build an efficient processing logic. In this direction, a traditional way is to use Digital Signal Processor (DSP) or Field Programmable Logic Array (FPGA) devices to achieve real time result calculation, and implement the calibration and correction functionality in software [9][10], or use oscilloscopes to study the detectors[11]. In our work, we aim to integrate all the signal processing and correction logic within one FPGA device, in order to simplify the system structure and guarantee a good flexibility. Meanwhile, the FPGA should also contain the function of online histogram construction in calibration.

This paper presents the design of a real-time digital signal processing logic in the Singles Processing Unit (SPU) electronics for a marmoset brain PET system. The logic is one key part in the SPU. It receives the amplitude and time digitization results and is responsible for all the necessary calculation and correction processing, which provide the position, time, and energy information for image construction, as mentioned above. Furthermore, functionality of flood map and energy spectrum construction is also included, which facilitates the calibration of the whole PET system. Considering the complexity and resource consumption of the correction LUTs, flood maps, and energy spectrum, we also propose methods to address issues and integrate all the above functions

Manuscript received June 24, 2018; revised March 25, 2019.
This work was supported by CAS Center for Excellence in Particle Physics (CCEPP). The authors are with the State Key Laboratory of Particle Detection and Electronics, University of Science and Technology of China, Hefei, 230026; and Modern Physics Department, University of Science and Technology of China, Hefei, 230026, China (telephone: 086-551-63607746, corresponding author: Lei Zhao, e-mail: zlei@ustc.edu.cn).
© 2019 IEEE. Accepted version for publication by IEEE. Digital Object Identifier 10.1109/TNS.2019.2908220

within one FPGA device.

This paper is organized as follows. Firstly, the structure of the PET system involved in this paper is introduced. Secondly, the structure of the SPU is presented. Thirdly, the digital signal processing and correction logic is discussed in detail. Finally, the results of both electronics performance tests and initial commissioning tests with the detector are presented.

## II. THE STRUCTURE OF THE PET SYSTEM

The block diagram of this small animal PET system is shown in Fig.1.

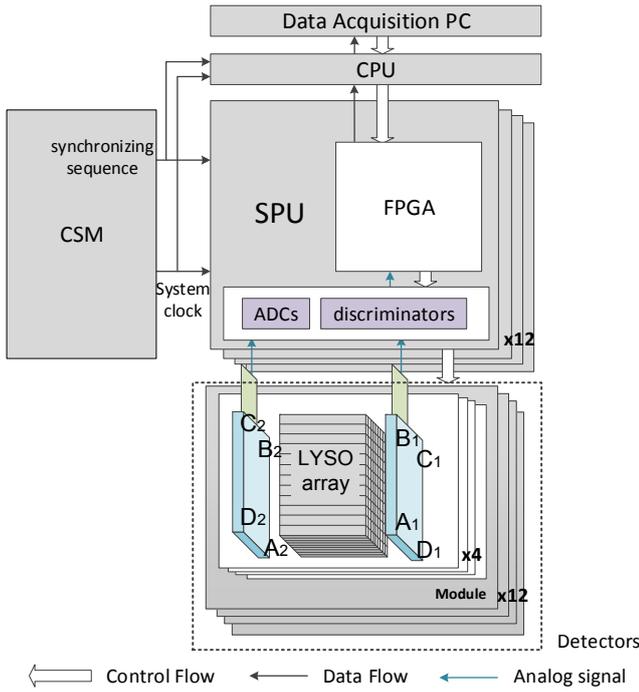

Fig. 1. Block diagram of the small animal PET system.

The detector consists of four 55 mm radius detector rings. In each ring, twelve 23×23 1mm×1mm×20 mm lutetium-yttrium oxyorthosilicate (LYSO) crystal [12][13] arrays are integrated, a total of 48 LYSO crystal arrays are included. As for each crystal array, there are two layers of 8×8 silicon photomultipliers (SiPM) [14][15] with an effective area of 25.6mm×25.6mm placed on the two opposite ends of the array and on each end, the resistive network is utilized for the signal readout of the SiPMs. From this dual-end detector structure the Depth Of Interaction (DOI) information [16][17] can be calculated, which would guarantee a higher spatial resolution for the image reconstruction [18][19].

As for the electronics, the signals from the pre-amplifiers close to the detectors are fed to the SPUs. Each SPU processes the output signals of one detector module, which contains four detector blocks (One detector block includes the aforementioned one crystal array and the two layers of SiPMs). The SPUs calculate all the needed information (i.e. position, time, and energy information) through the logic in the Field-Programmable Gate Array (FPGA), and then transfer the data to a Coincidence Processing Unit (CPU), which sorts the data packages according to the time information, seeks the data belonging to one event (corresponding to two 511 keV photons emitted in one annihilation), and finally creates Lines Of Response (LORs). The results of CPU are then collected by the remote Personal Computer (PC) for image reconstruction. The SPU also receives a system clock and synchronizing sequences from the Clock and Synchronization Module (CSM) for synchronization among the 12 SPUs.

## III. THE STRUCTURE OF THE SPU

As shown in Fig.2, in each block, eight analog output signals (marked as $A_1$, $B_1$, $C_1$ and $D_1$ for one end of the detector, and $A_2$, $B_2$, $C_2$ and $D_2$ for the other end) from the pre-amplifiers of the dual-end detector are transmitted to one SPU.

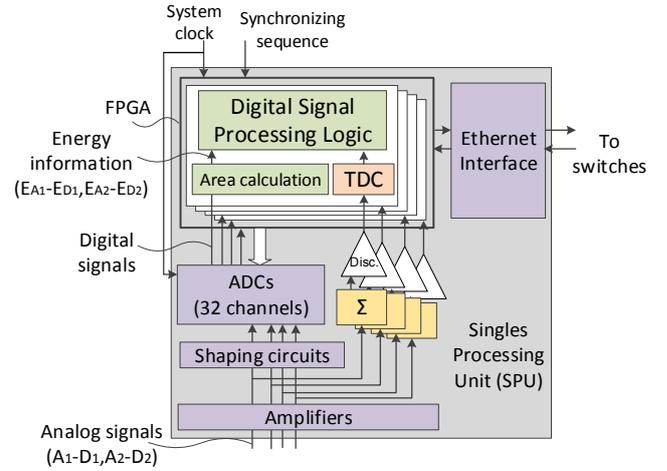

Fig. 2. Block diagram of the SPU.

With regard to the position and energy measurement for each block, the eight input signals are further amplified, and pass through shaping circuits before being fed to the 12-bit 62.5-MHz sampling rate AD9222 Analog-to-Digital Converters (ADCs) from Analog Device Inc. The digitized waveforms from the ADCs are then transferred to a Xilinx Artix-7 FPGA [20] for area integral calculation to obtain the energy information of the eight channels (marked as $E_{A1}$, $E_{B1}$, $E_{C1}$, and $E_{D1}$ for one end of the detector, and $E_{A2}$, $E_{B2}$, $E_{C2}$, and $E_{D2}$ for the other end, as shown in Fig. 2). Then, these eight energy values are transferred to the digital signal processing logic for further processing. As for the time measurement, the eight input signals are first amplified, then fed to an analog summation circuit, and finally discriminated and digitized by the Time-to-Digital Converter (TDC) in the FPGA. The implemented TDC is based on multiphase clock interpolation method [21][22] with a bin size of 333 ps [21]. The time measurement results are also fed to the digital signal processing logic. As mentioned above, the final data results are collected through switches by the CPU for coincidence and finally transferred to the remote PC. Meanwhile, the SPU receives the commands and configuration data (such as correction LUTs) from the PC through the switches.

## IV. DESIGN OF SPU DIGITAL SIGNAL PROCESSING LOGIC

To guarantee the flexibility of the SPU electronics, the digital processing logic integrates four processing modes within one FPGA, which is selected conveniently by the command issued by the PC. The four processing modes include the regular data processing mode, flood map construction mode, energy spectrum construction mode, and raw data mode. Fig. 3 shows the structure of the digital signal processing logic.

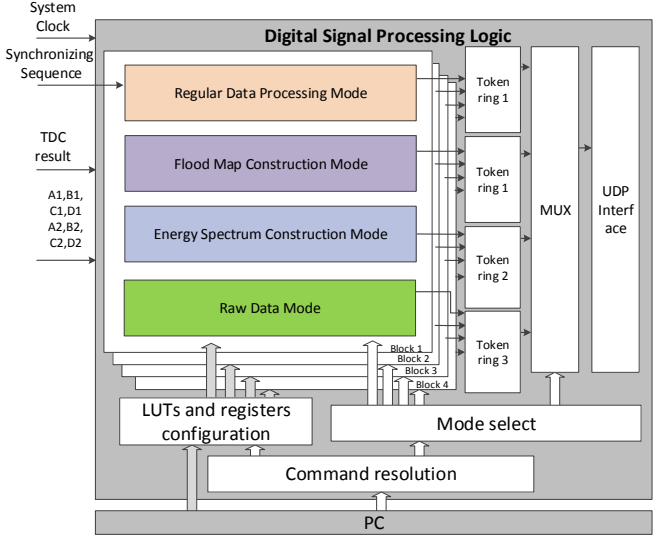

Fig. 3. Block diagram of the SPU digital signal processing logic.

Regular data processing mode, which calculates the position, time and energy information, is mainly used in common imaging process of the small animal PET system. Meanwhile, the flood map construction mode and energy spectrum construction mode are mainly used to obtain Crystal Look-up Table (CLT) and photon peak LUT. These two modes are used for system calibration. The raw data mode is also integrated in the logic to facilitate the system debugging.

During the logic design, much attention is paid to the issue of resource consumption. On one hand, there exist a series of LUTs in all the processing modes except for the flood map histogram mode, among which the CLT consumes large amount of resources. On another hand, in the two statistic modes, big memories are also required. As for each block (4 blocks in each SPU), the flood map is a matrix of $512 \times 512$, each cell with a 10-bit data width, and the energy spectrum consists of $23 \times 23$ Random Access Memories (RAMs), each with a depth of 256 and data width of 10 bits. The CLT, the flood map and the energy spectrum construction consume the majority of the memory resources.

Although large memory can be implemented by using an external RAM, it would cause higher system complexity. Besides, different real-time digital signal processing blocks would share one common data interface of the external RAM, and this means that the data streams would need to be switched through complex routes for which timing closure is more difficult to obtain. Therefore, we devoted efforts to reduce the logic consumption, and proposed methods to implement the above memories.

The detailed information of the different processing modes and the methods to reduce the logic consumption are presented in the following sub-sections.

### A. Regular Data Processing Mode

In actual application, the regular data processing mode calculates the reaction crystal address (i.e. position information), energy, and time measurement result of each event. The block diagram of the signal processing in this mode is shown in Fig. 4.

To obtain the position results, we first calculate raw position of each event in X and Y directions (marked as raw (X, Y)) with a bit width of 18 bits (9 bits for X and 9 bits for Y) and the 4-bit DOI information from the eight energy values through the center-of-gravity method, as in:

$$X = 0.5 \times \left( \frac{E_{A1} + E_{D1}}{E_{A1} + E_{B1} + E_{C1} + E_{D1}} + \frac{E_{A2} + E_{D2}}{E_{A2} + E_{B2} + E_{C2} + E_{D2}} \right) \quad (1)$$

$$Y = 0.5 \times \left( \frac{E_{A1} + E_{B1}}{E_{A1} + E_{B1} + E_{C1} + E_{D1}} + \frac{E_{C2} + E_{D2}}{E_{A2} + E_{B2} + E_{C2} + E_{D2}} \right) \quad (2)$$

$$DOI = \frac{E_{A1} + E_{B1} + E_{C1} + E_{D1}}{E_{A1} + E_{B1} + E_{C1} + E_{D1} + E_{A2} + E_{B2} + E_{C2} + E_{D2}} \quad (3)$$

With the raw (X, Y), we can finally identify the interaction crystal location through the CLT.

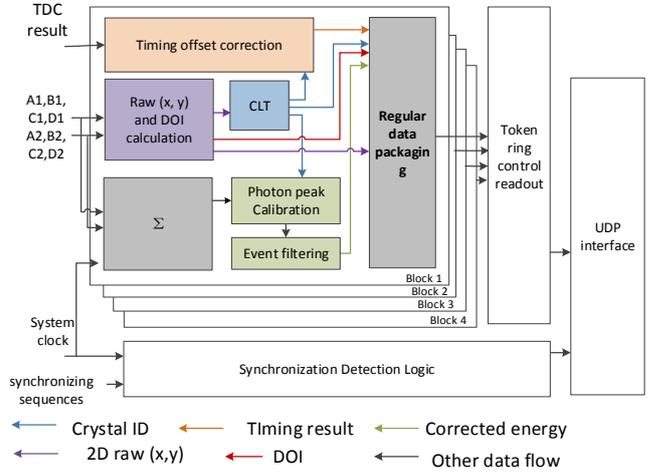

Fig. 4. Block diagram of regular data processing mode.

As for energy measurement, the eight energy values are summed together, and the result is then converted to the energy result with the unit of keV, by multiplying a coefficient. Considering the mismatch due to the crystals, SiPMs, and the electronics, the coefficients vary with different crystals. These coefficients are stored in the photon peak LUT (established through the result of the energy spectrum construction mode, which will be discussed later), and are used for the energy correction. After that, the corrected results are filtered by an energy window with an adjustable range, in order to reject

Compton events.

Time correction is also required considering the delay mismatch caused by the crystals, SiPMs, and electronics. In this correction, we first locate the crystal address according to the position measurement results, and then read a time offset value from the corresponding LUT, which contains the delay information crystal by crystal. By subtracting this offset from the output of the TDC, we can finally obtain the time results.

The final position, energy and time results within one block are packaged together as one regular data frame, and then buffered in the block FIFOs. As for the data readout, a token ring structure is applied. The maximum required data rate of the logic is 1,000,000 events/s/block as mentioned before, which means only one regular data frame is buffered in each block FIFO in 1 μs at most. The token ring is executed under a 125 MHz clock and it takes only 4 clock cycles (i.e. 32 ns) to transfer all the data in the block FIFO to the module FIFO. This guarantees that the regular data frames stored in the module FIFO are sorted automatically according to their time information. The data frames within a pre-determined time period are further packaged by adding headers, which are transferred to the CPU for further data package sorting as mentioned above.

As aforementioned, the data from 12 SPUs need to be synchronized. To monitor the status of the SPUs, each SPU receives the system clock and a synchronizing sequence. The SPU compares the output of the internal time stamp counter and the time stamp value contained in the synchronizing sequence to judge whether the system is in a correct condition. If not, a special synchronization data package will be inserted into the regular data package flow to report to the PC and reset the internal time counter with the received value.

Fig. 5. Shows the timing diagram of the regular data processing logic utilized in the each block, which is a complete pipelined logic. Firstly, the calculation of the raw (x, y) and DOI information from the eight energy values costs $K$ cycles of the 125-MHz clock. $K$ is an adjustable parameter when implementing the pipelined Xilinx IP core -- Divider Generator to perform the division calculation. Secondly, the subsequent "crystal locating" process costs two 125-MHz clock cycles to search the CLT and one cycle to compare the result with the raw (x, y) value to obtain the crystal address. Finally, the simultaneously performed energy and the time information correction cost 2 cycles to search the LUT and 1 cycle to calculate. Since it is a pipelined structure, it only introduces a total latency of ($K$+3+3) 125-MHz clock cycles and no extra dead time. Therefore, the working speed of this logic is sufficiently high for the application.

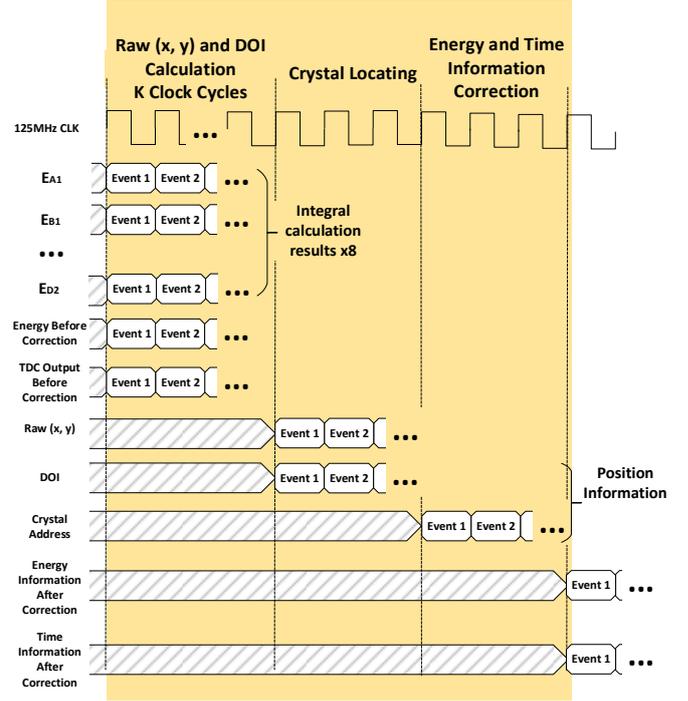

Fig. 5. Timing diagram of the regular data processing mode.

### B. Flood Map Construction Mode and Energy Spectrum Construction Mode

A flood Map is a two Dimensional 512×512 histogram calculated from a large amount of raw (X, Y) position data. To improve the efficiency of the histogram construction process, the online construction mode, as shown in Fig.5, is integrated in the FPGA.

As shown in Fig. 6, the raw (X, Y) is used as an 18-bit address to readout the corresponding count value from the RAM and then add the value by 1 if no overflow occurs. When the data volume is enough, the content of the RAM is read out as the final flood map result, which is used to establish the CLT using software.

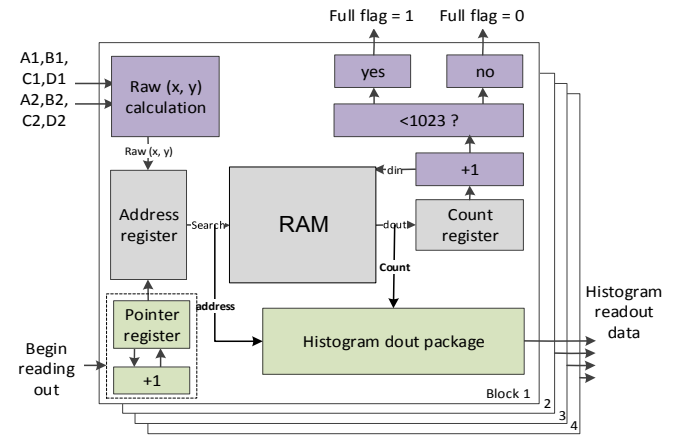

Fig. 6. Block diagram of the on-line flood map construction mode.

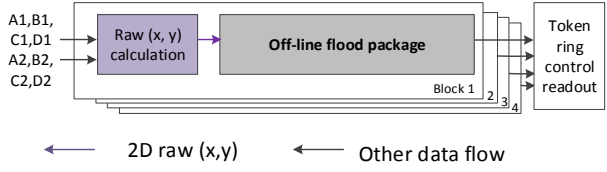

Fig. 7. Block diagram of the off-line flood construction mode.

To ensure the flexibility of the system, an offline construction mode is also supported. As shown in Fig. 7, the raw (X, Y) is directly read to PC, and a flood map can be built based on software programming.

The energy spectrum is used to obtain the photon peak LUT. In this construction mode, we need to access 529 memory segments with 256 address locations each and 10-bit data width for the 23×23 crystals in each detector block. Similar to the above construction process, both online and offline construction processes are supported.

As mentioned above, one of the largest memory resource consumption occurs in these two construction modes. Considering that the two construction processes are not required to function simultaneously and the RAMs used in these two modes have the same data width, we implement one 10-bit-width 512×512-depth block RAM in each block, and share it in the two modes using multiplexers, according to the configuration commands from the PC.

### C. Raw data mode

In this mode, the energy for each channel and time measurement results for each block are directly packaged and transferred out. This mode allows the users to observe each channel individually. With the output data in this mode, the user can also perform all necessary analysis based on software programming on the remote PC.

### D. Design of boundary CLT

As mentioned above, the other largest memory resource consumption is the CLT. To reduce the memory size, we propose a structure named the boundary CLT.

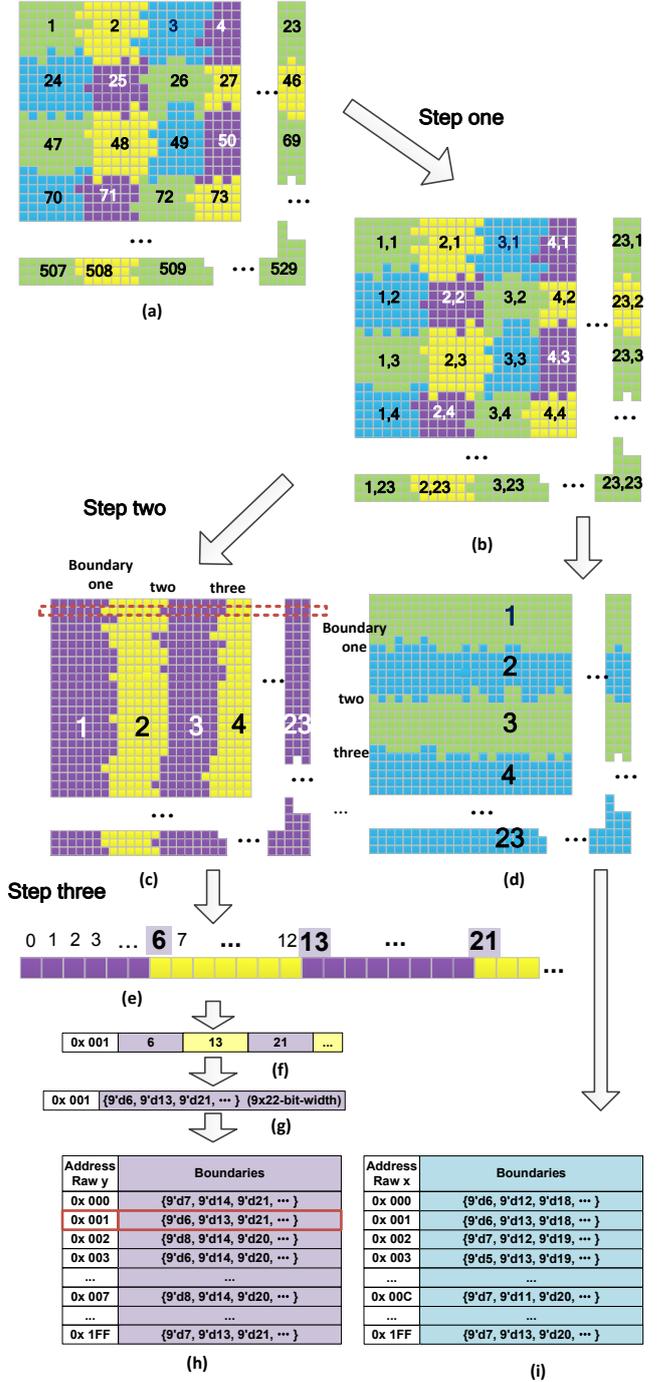

Fig. 8. Transformation from a traditional CLT to a boundary CLT.

Fig.8 (a) shows a traditional structure to build the CLT. As mentioned above, one detector block is divided to 512 × 512 pixels in a large square, as shown in Fig. 8 (a), and multiple pixels would correspond to the same crystal, for instance, the pixels in green color in the left-top range of the square correspond to crystal No. 1. In the traditional way, each pixel is implemented as one 10-bit wide memory cell, all the pixels of the same crystal would store the same number, ranging from 1 to 529. When a new raw (X, Y) position is calculated, it is used as the address to search the RAM, and read out the content of the memory cell (i.e. crystal address). In such structure, it consumes a memory space of 512 × 512 × 10 bits=2.5 Mb. In fact, with higher position resolution of raw (X, Y), much more memory is required. For example, if one block is divided to $2^n \times 2^n$ (n is an integer, and in our PET system, n=9, corresponding to 512 × 512), corresponding to a total crystal number of $k$, the total resource consumption can be estimated as

$$Memory\ Size = 2^n \times 2^n \times \lceil \log_2 k \rceil = 2^{2n} \times \lceil \log_2 k \rceil \quad (4),$$

Where $\lceil \log_2 k \rceil$ represents the smallest integer larger than the real number $\log_2 k$.

We observe the structure in Fig. 8 (a), and find that redundant information is contained in the RAM. Therefore, our idea is to shrink the memory by reducing the redundancy. We transform the traditional CLT to a boundary CLT as shown in Fig.8. The traditional CLT is resolved into two boundary CLTs in two directions. Three transformation steps are performed as follows: firstly, the crystal address encoding is transformed to two dimensional coordinate, for example, Crystal No. 24 in Fig. 8 (a) to Crystal No. (1, 2) in Fig. 8 (b). Secondly, Fig. 8 (b) is resolved into Fig.8 (c) and Fig. (d), and the related regions are combined and re-categorized. For example, Crystal No. (1, 1), (1, 2) … to (1, 23) is re-categorized into Region No. 1 in Fig. 8 (c). In the third step, we can finally establish the boundary CLT and store the content of Fig. 8(c) into a RAM (Fig. 8(h)). Each boundary CLT has 22 boundaries. Each boundary actually consists 512 tiny sections, and can thus be stored in a 512-depth memory. Considering the boundary ranges from 1 to 512, only 9 bit data width is enough. With such a structure, we only need a memory space of 512×22×9 bits=0.099 Mb. This means the resource consumption is successfully reduced from 2.5 Mb to 0.099 ×2=0.198 Mb (Fig. 8 (c) plus Fig. 8 (d)). Using the definition of n and k in (4), the memory size with this structure can be expressed as

$$Memory\ Size = n \times (\sqrt{k} - 1) \times 2^{n+1} \quad (5),$$

from which we can observe the resource consumption is greatly reduced.

In real application, we implement the 512×22×9 bit memory to a 512×198 bit RAM (as shown in Fig. 8 (h) and Fig. 8 (i)), considering the minimum bit width occupation of the RAM in real implementation.

With the above boundary CLT, the raw (X, Y) are directly compared with the boundaries stored in the two CLTs corresponding to Fig. 8 (c) and (d). We use Y information as the address to readout one column of CLT as in Fig. 8(c), and then compare the X information with the 22 boundaries, and directly locate one dimension of the crystal number. With the same procedure using the other CLT, we can finally obtain the two dimensional crystal address. The last step is to transform this crystal address back to the one dimensional range of 1 to 529. The detailed process is shown in Fig. 9.

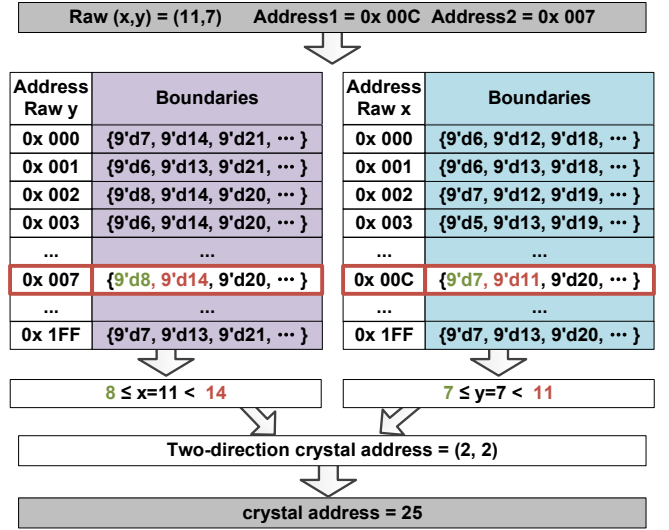

Fig. 9. Crystal identification with the boundary CLTs.

With the above methods to optimize the memory structure, we can significantly reduce the memory consumption in the FPGA. TABLE I shows the consumption before and after optimization, and now only a total of 10.89 Mb is required to implement all the needed RAMs.

TABLE I
MEMORY RESOURCE CONSUMPTION

| LUTs, flood map and energy spectrum | Resource occupation (theoretically, 4 block together) / Mb | |
|---|---|---|
| | before optimization | after optimization |
| Boundary CLTs | 10 | 0.76 |
| Time correction LUTs | 0.07 | 0.07 |
| Photon peak LUTs | 0.06 | 0.06 |
| Flood map and Energy spectrum | 10 and 5.16 | 10 |
| Total | 25.29 | 10.89 |

### E. Data Transfer Interface

We estimate the data rate of one SPU before data interface design. The probability of gamma photons hitting a detector block in this small animal PET is

$$Probability = \frac{Area}{4\pi \times r^2} \approx \frac{(25.6\,\text{mm})^2}{4\pi \times (55\,\text{mm})^2} = 1.72\% \quad (6),$$

where *Area* refers to the effective area of a detector block and *r* refers to the radius of the detector ring, as mentioned in Section II.

In real application, a maximum radioactivity is 200 μCi (Curie), i.e., 7.4 MBq (14.8 M singles/s) and the detection efficiency of LYSO crystals on gamma photons is 80%. The average events for one SPU is 14.8M×1.72%×80%×4 ≈0.82M. According to the data package frame, there are 16 bytes in each event, the total average data rate is 0.82 M×16 byte = 105 Mbps.

Gigabit Ethernet interface based on the User Datagram Protocol (UDP) [23] is employed for data transfer. The transport layer and the network layer is implemented through logic design, while an intellectual property core is employed as the link layer and an 88E1111 chip [24] of Marvell Inc. is used as the physical layer.

*F. Resource Consumption of the whole logic*

The total resource consumption of the whole logic in the FPGA is listed in Table II.

TABLE II
TOTAL LOGIC CONSUMPTION OF THE WHOLE LOGIC

| Resource | Utilization | Available | Utilization % |
|---|---|---|---|
| LUT | 37783 | 133800 | 28.24 |
| LUTRAM | 1573 | 46200 | 3.40 |
| FF | 33236 | 267600 | 12.42 |
| BRAM | 313 | 365 | 85.75 |
| DSP | 5 | 740 | 0.68 |
| IO | 163 | 500 | 32.60 |
| BUFG | 18 | 32 | 56.25 |
| PLL | 2 | 10 | 20.00 |

## V. TESTING RESULTS

*A. Performance Test Results of the SPU*

The performances of the SPU module include position, time, and energy measurement resolution. To evaluate the quality of the electronics, we conducted a series of tests on the SPU.

The system under test is shown in Fig. 10. We capture the output waveform of the aforementioned pre-amplifiers close to detectors with an oscilloscope. According to this waveform, we use the signal generator AFG3252 [25] from Tektronix Inc. to generate the input signal for the tests. This signal is fed to an attenuator for amplitude adjustment, and then serves as the input signal of the SPU. The data results of the SPU are collected and analyzed by a PC.

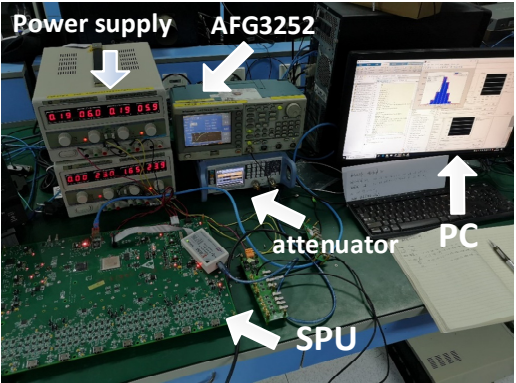

Fig. 10. System under test.

*1) Position Measurement Resolution*

In the regular data processing mode, we use the raw (X, Y) position to locate the corresponding crystal address. Therefore, a good resolution of the raw position measurement is required for the SPU.

We conducted tests in the flood map construction mode. We changed the amplitude ratio among the eight input channels, and simulated different interaction positions. We conducted tests on nine positions, covering the center, corners and edges, to evaluate the SPU performance.

Fig.11 shows the bar graphs of the X and Y position testing results on all the nine positions of one detector block. Fig.12 shows the bar graphs of X and Y position testing results of one SPU module, which indicate that the position resolution is better than 1.5‰ RMS for all the blocks.

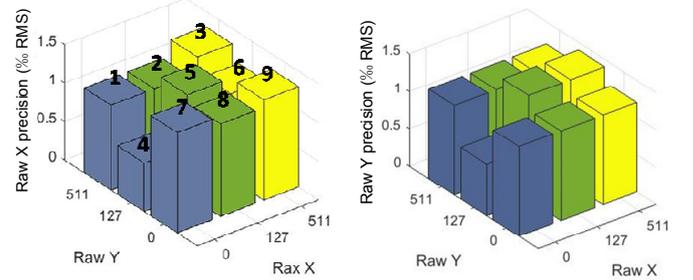

Fig. 11. The position resolution test results of one block (nine positions).

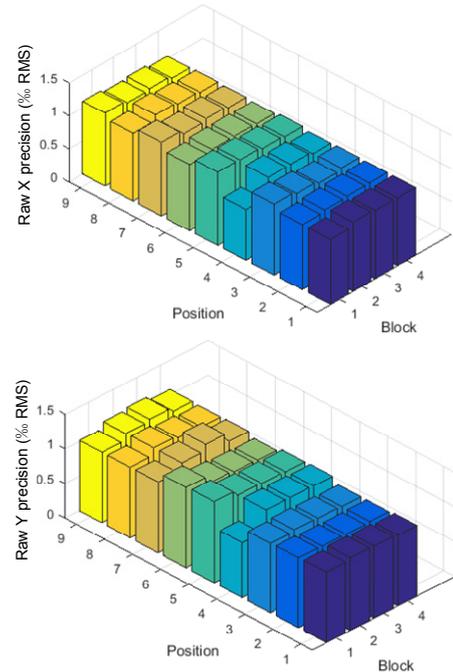

Fig. 12. Position resolution test results of one SPU module.

*2) Time Measurement Resolution*

The "cable delay test" method [26] is employed to evaluate the time measurement resolution.

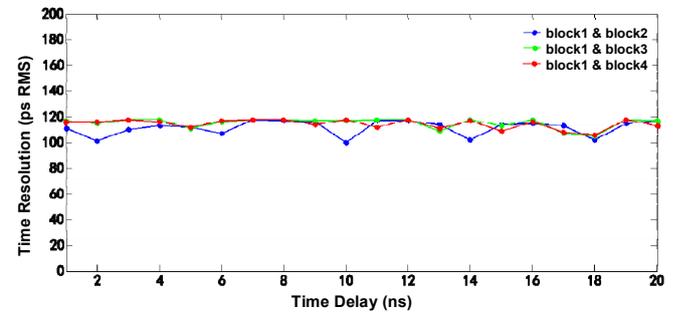

Fig. 13. Time resolution test result of each channel.

As the 8 channels are summed before discrimination, there are one time measurement channel for each block. In the cable delay test method, we used the AFG3252 to output two synchronized pulse signals which are fed to different blocks, and calculated the RMS value of the time difference between two blocks (block 1 & block 2, block 1 & block 3 and block 1 & block 4). Considering that the two time measurement channels are not interrelated, we can get the time resolution of single channel by dividing the above RMS value by $2^{1/2}$. We changed the time difference between the two input signals, and conducted a series of tests. The results are shown in Fig. 13, from which we can observe the time resolution of each channel is better than 120 ps RMS.

*3) Energy Measurement Resolution*

Fig.14 shows a typical energy testing result of one channel of an SPU module at 2.0 V input. We can calculate the normalized energy resolution (the FWHM divided by the mean value), and the result is 2.4‰ FWHM.

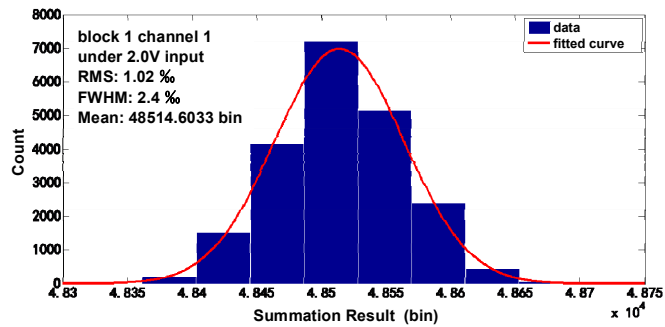

Fig. 14. Typical testing result of energy resolution at 2.0 V input.

Fig. 15 shows the energy resolution of all the 32 channels in one SPU module in the input amplitude range from 0.2 V to 3.0 V, and the energy resolution is better than 5‰ FWHM for all the channels.

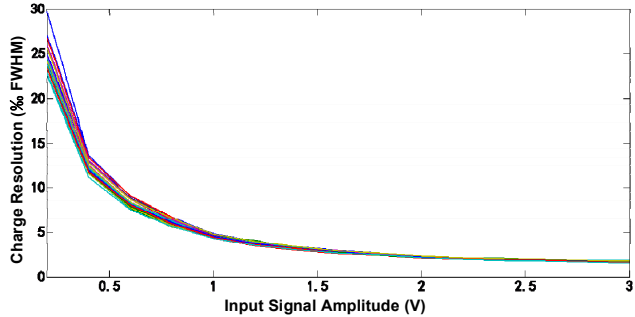

Fig. 15. Energy resolution of all 32 channels with different input signal amplitudes.

### B. Initial Commissioning Test with the Detectors

After above tests, we further conducted initial commissioning tests with the detectors using $^{22}$Na.

*1) The Flood Map and Energy Spectrum Construction*

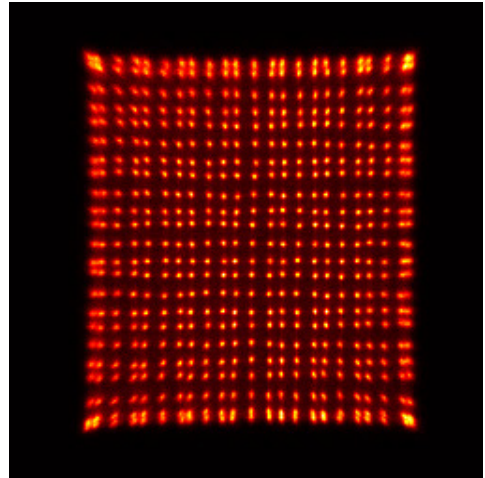

Fig. 16. The 512×512 flood map test results.

Fig. 16 shows the 512×512 flood map, which is constructed in the online mode. In this flood map, we can clearly observe 23×23 LYSO crystals in one block, which agrees well with the expected behavior.

Fig. 17 shows the online constructed energy spectra of two crystals (No. 76 and No. 79) before photon peak correction. We observe Fig. 17, the peaks of these two crystals are at different places before the correction.

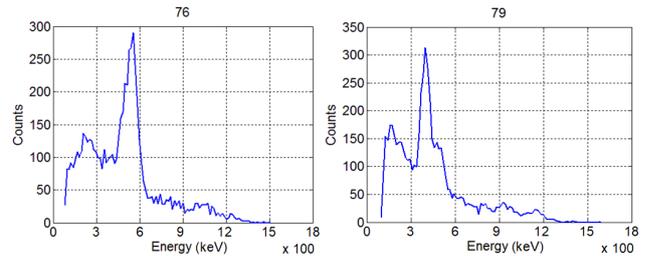

Fig. 17. Energy spectrum of two crystals before photon peak correction.

*2) Photon Peak Correction to 511 keV*

Fig. 18 shows the energy spectra of the two corresponding crystals after correction using the photon peak LUT obtained in the energy spectrum construction mode in a different acquisition process. The results indicate that both of the peaks are corrected to 511 keV, which proves the logic functions well.

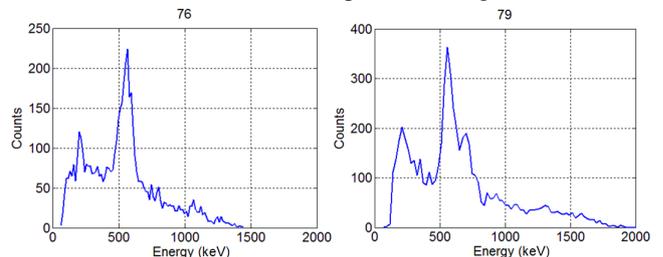

Fig. 18. Energy spectrum of the two crystals after photon peak correction.

## VI. CONCLUSION

An FPGA based real-time digital signal processing and correction logic is designed for a small animal PET. The results of both the electronics performance tests and the initial commissioning tests with the detector indicate that the logic design meets the application requirement.

ACKNOWLEDGMENT

The authors would like to thank Dr. Yongfeng Yang, Yibao Wu, Xiaohui Wang, Ziru Sang and Zhonghua Kuang in ShenZhen Institutes of Advanced Technology, CAS for their kind help.